\documentclass[]{raa}            
\usepackage{graphicx,times}
\usepackage{natbib}

\providecommand{\bm}[1]{\mbox{\boldmath$#1$\unboldmath}}

\begin{document}

\title{The nature of the companion of PSR J1719-1438:\\ a white dwarf or an exotic object?}

 \volnopage{ {\bf } Vol.\ {\bf } No. {\bf }, 000--000}
   \setcounter{page}{1}

   \author{J. E. Horvath
      \inst{1}
   }

\institute{Instituto de Astronomia, Geof\'\i sica e Ci\^encias
Atmosf\'ericas - Universidade de S\~ao Paulo, Rua do Mat\~ao,
1226, 05508-900, Cidade Universit\'aria, S\~ao Paulo SP, Brazil}
{\it foton@astro.iag.usp.br}
\date
\vs \no
   {\small Received [year] [month] [day]; accepted [year] [month] [day]
   }

\abstract{We rise in this {\it Letter} the possibility that the very dense, compact
companion of PSR J1719-1438, having a Jupiter-like mass is an exotic
quark object rather than a light helium or carbon white dwarf. The exotic hypothesis
explains naturally some of the observed features, and gives quite
strong predictions for this system, to be confirmed or refuted in
future feasible studies.}

 $\cdots\cdots$ \keywords{Binary pulsars ; exotic matter }

   \authorrunning{J.E. Horvath }            
   \titlerunning{Exotic nature of the companion of PSR J1719-1438}  
   \maketitle


%
%
\section{Introduction}           
\label{sect:intro}

The presence of bodies orbiting pulsars (planets) was quite unexpected
before their discovery by \cite{WF92}, and prompted an ample discussion about
their formation scenarios \citep{Phi93}. Overall, the fraction of pulsars
with planets does not now seem very large, suggesting special conditions for
their formation, rather than a generic channel yielding a large number of
pulsar-planet systems. A recent work \citep{B11}, however, added an important twist
to this problem: the detection of a Jupiter-mass object around the pulsar
PSR J1719-1438, a 5.7 ms pulsar, has been interpreted as a case in which
an Ultra Low-Mass X-ray Binary suffers a transformation of a white dwarf
into a planet. Moreover, this downsizing suggests, to comply with the condition
of the companion radius fitting inside the Roche Lobe (estimated to be
smaller than $4.2 \times 10^{4} km$ for the most favorable parameters),
a heavier-than-helium composition of the latter. An important corollary of the study
is the existence of a {\it minimum density} for the companion object, a direct
consequence of the orbital period and Roche Lobe considerations \citep{B11},
namely

\begin{equation}\label{eqn:1}
\rho = {3 \pi \over{(0.462)^{3} G P^{2}}} \, \geq 23 \, g \, cm^{-3}
\end{equation}

This average is far in excess of the density of normal
Jupiter-sized planets and prompted the evaporated helium/carbon
white dwarf picture (in fact an helium object would be contrived,
but not impossible because of size constraints \citep{BVH}).

Our main argument is that the existence of the lower bound eq.(1)
allows an alternative interpretation in term of exotic matter,
namely versions of stable or metastable quark matter, discussed
over the last decades \citep{Wit84, AFO86, GKW95, Mich88}. As
suggested below, this is a sound alternative, even if not yet
compelling, and would have, if confirmed, deep implications for
the nature of dense matter. We point out and discuss in the
remaining of this {\it Letter} the sketch and a few direct
consequences of this hypothesis.

\section{Variants of quark matter composition for the companion object}

The simplest possible exotic object candidate for PSR J1719-1438 is a
planetary mass, superdense strange quark matter nugget \citep{Wit84, AFO86}
These chunks of cold strange quark matter (SQM) can exist all the way down to
microscopic masses (then referred as {\it strangelets}) by hypothesis, with
densities $\sim \, 4B \, \sim \, 4 \times 10^{14} g cm^{-3}$ in the limit in which
gravitational forces are not important. In fact, a nugget with the mass of
$\sim$ Jupiter could {\it not} have formed in the early Universe because of the horizon
constraints \cite{Wit84}, even if the SQM hypothesis is true. However, the
ejection of a planetary mass nugget is quite possible in astrophysical
events (see below). The radius of a Jupiter mass nugget is just
$\approx 1 km$, and satisfies all observational constraints easily.

While the SQM nugget is homogeneous by construction, the possible
existence of {\it structured}, symmetric quark states at low
baryon number $A$ has been considered in the past, leading to a
different type of matter in macroscopic astrophysical objects. The
calculations of the so-called H dibaryon (a quark analogue to a
$\Lambda-\Lambda$ state) led to the suggestion of its
(meta)stability \citep{J77}, recently reinforced by lattice QCD
calculations \citep{Ino11}, although it is yet to be found in
experiments. A novel state, the H-matter \citep{Tama91},
conceptually analogous to neutron matter, would appear, which is
especially interesting in the case of a stable dihyperon but also
relevant for the metastable case. It is interesting to note that
this proposal has been recently revived by Lai, Gao and Xu (2011),
and by Occam's razor points toward a common composition for the
pulsar and its companion. A still bolder proposal is the
completely symmetric 18-quark state, termed originally as the
$Q_{\alpha}$ because of the suggested analogy with the helium
nucleus \citep{Mich88}. There is no experimental evidence for its
presence, and only very rough estimations of the binding energy of
this state $(uds)^{6}$ with $A=6$ and $S=-6$ are available, which
may be stable even if the $H$ particle is {\it not} bound with
respect to the $\Lambda\Lambda$ threshold \citep{STY11}.

The structure of macroscopic pure H-matter and $Q_{\alpha}$
objects has not been fully addressed (however, see \cite{BHV90}
for their possible role of the latter in compact stars). Since
$Q_{\alpha}$s are a spin-0 uncharged state, it is quite natural to
treat them as bosons with a repulsive (hardcore-like) interaction.
A phenomenological description of self-interacting bosons applied
to stellar structure has been first discussed in \cite{CSW86}.
Further work \citep{DS88} derived an effective equation of state
in the low-density and high density limits specific to a scalar
diquark with mass $\sim 575 MeV$. These results can be readily
adapted for the symmetric strangelet case, like the 18-quark
$Q_{\alpha}$ \citep{Mich88}) by appropriately changing the mass
and coupling constant values, giving a maximum mass for a stellar
model of

\begin{equation}\label{eqn:2}
M_{max} = \, 0.03 \,\,  {\biggl({\lambda \over{25}}\biggr)}^{1/2} \,\,
{\biggl({6 GeV \over{m_{A}}}\biggr)}^{2} \,\, M_{\odot}
\end{equation}

This mass is comfortably within the required range unless the mass
of the lightest bosonic strangelet is much heavier, but rather
insensitive to the exact value of the self-interaction parameter
$\lambda$, which remains poorly determined. However, the radius is
still orders of magnitude smaller than the ones calculated in the
recent proposal of strangelet dwarfs by \cite{AHR11}, which is
determined by electrostatic forces. Because of the charge of the
considered strangelets, the latter is in the ballpark of more
``normal'' white dwarfs, and quite insensitive to the exact mass.
The same happens with the former proposal of strange dwarfs by
\cite{GKW95}. The point of these rough estimates is to remark that
an exotic Jupiter-mass object could be very different in size,
truly point-like indeed, and thus easier to fit within the
observed features.

\section{Formation mechanisms for the system}

It is clear from the outset that an exotic nature of the companion would force
to consider a strange star nature for the pulsar itself \citep{BHV90, JRS06},
and therefore a common origin of both objects. Among the possible common origin
scenarios, we should remark that the so-called fallback model \citep{Phi93}
for the formation of planets around pulsars must be {\it extended}
in this case to consider material
ejected by the formation of the compact star itself. This leads back to the idea
of a SQM driven explosion \citep{BH89}, in which the conversion
$n \rightarrow uds + energy$ process is inevitably turbulent \citep{HB88,
BH89, H10}, a proposal recently confirmed by direct numerical calculations \citep{HR11},
setting the stage for the ejection of exotic matter as an effect of large turbulent,
high-velocity eddies. The companion would form out of this ejected matter, as a
result of high rotation.

Another possibility for the formation of the PSR J1719-1438 system
is the merging of two low-mass strange stars, recently shown
\citep{HT09} to eject up to $0.03 M_{\odot}$ or an order of
magnitude more matter than needed, which would be acceptable
provided the sum of both initial masses does not exceed the
limiting mass of strange stars sequence (note that a naive balance
of tidal torques and surface tension of one nugget had predicted a
mass of $\sim 10^{-18} M_{\odot}$ \citep{Jes01}, thus an enormous
number of asteroid-size nuggets instead).

Last, but not least, we would like to point out that a third
scenario, the merging of two white dwarfs \citep{Phi93}, leading
to the so-called Accretion-Induced-Collapse (AIC) forming a single
``neutron'' star should necessarily be revisited concerning the
formation of SQM. This is because the microscopic conditions for
SQM formation would be achieved anyway when supranuclear densities
are achieved and, therefore, in addition to be an attractive
channel for the direct formation of a millisecond pulsar, the
scenario would have the bonus feature of providing some {\it
exotic} ejected matter as well, yet to be calculated and
quantified. Indeed, it is quite unlikely to make a neutron star
without also converting it to a strange star in the same AIC
process.

\section{Conclusions}

We have argued in this {\it Letter} that an exotic object is
viable as a companion of the millisecond pulsar PSR J1719-1438,
and in fact a unique alternative to a helium/carbon WD. To
distinguish both, there are a few theoretical scenarios to be
explored in detail, and some key observations worth performing. A
similar idea (namely, an exotic composition) was formerly advanced
by Xu and Wu (2003) for the case of Wolczan's planets, for which
less information is available even today. If the companion of PSR
J1719-1438 is an exotic object, it would easily explain why there
is no observed eclipses in spite of a (nearly) edge-on
inclination. Actually, the strong photometric limits derived by
\cite{B11} using Keck-LRIS instrument can not be used to place
constraints to the inclination either. The cases of a SQM nugget
or structured strangelet the size would be too small to detect any
photometric signal, while if a strangelet dwarf is realized, its
size would be $\sim 5 \times 10^{3} km$ \citep{AHR11}, but the
surface properties still depend on the existence or absence of a
normal matter atmosphere to reprocess the pulsar radiation. The
strange white dwarfs of \cite{GKW95} would be difficult to
distinguish from conventional carbon/helium ones.

The exotic nugget model predicts that no (carbon or helium) lines
should be ever observed associated to the companion, in contrast
to the expectations for a normal white dwarf or exotic analogue
counterparts \cite{AHR11, GKW95}, which should show its nature at
some level of sensitivity. In addition, the lack of detection of
evaporation signatures would be naturally accommodated, since no
evidence of circumstellar material is found \citep{B11}. Finally,
and because of angular momentum considerations of an ejected
nugget, we expect that the orbit $\overrightarrow{L}$ to be
aligned with the spin of the pulsar. These are quite
straightforward predictions and unique to be checked in futures
studies addressing the nature of this system.

\normalem
\begin{acknowledgements}
We acknowledge the financial support received from the Funda\c
c\~ao de Amparo \`a Pesquisa do Estado de S\~ao Paulo. J.E.H.
wishes to acknowledge the CNPq Agency (Brazil) for partial
financial support.
\end{acknowledgements}

\label{lastpage}

\end{document}